\documentclass[referee]{aa} % for a referee version
\usepackage{graphicx}
\usepackage{txfonts}
\usepackage{natbib}
\begin{document}

\def\dexkpc{$\textrm {dex kpc}^{-1}${}}
\def\deg{\ensuremath{^\circ}}
\defcitealias{Lem07}{Paper~I}
   \title{Galactic abundance gradients from Cepheids}

   \subtitle{On the iron abundance gradient around 10-12 kpc.}

   \author{B. Lemasle
          \inst{1,2}
          \and
           P. Fran\c cois
          \inst{2}
          \and
           A. Piersimoni\inst{3}
          \and
           S. Pedicelli\inst{4,5}
          \and
           G. Bono\inst{4,5}
          \and
           C. D. Laney\inst{6}
          \and 
           F. Primas\inst{4}
          \and 
           M. Romaniello\inst{4}\fnmsep
          \thanks{Based on observations obtained at the Canada-France-Hawaii Telescope (CFHT) which is operated by the National Research Council of Canada, the Institut National des Sciences de l'Univers of the Centre National de la Recherche Scientifique of France, and the University of Hawaii. Based on observations collected with FEROS at the European Southern Observatory (La Silla, Chile). Based on observations obtained at the Telescope Bernard Lyot (USR5026) operated by the Observatoire Midi-Pyr\'en\'ees and the Institut National des Science de l'Univers of the Centre National de la Recherche Scientifique of France.}
          }

   \offprints{B. Lemasle}

   \institute{Universit\'e de Picardie -  Jules Verne, Facult\'e des Sciences, 33 Rue Saint-Leu, 80039 Amiens Cedex 1, France\\
              \email{bertrand.lemasle@etud.u-picardie.fr}
         \and
              Observatoire de Paris, GEPI, 61, Avenue de l'Observatoire, 75014 Paris, France
         \and
              Istituto Nazionale di Astrofisica, Osservatorio Astronomico di Collurania, via M. Maggini, I-64100 Teramo, Italy
         \and
              European Southern Observatory (ESO), Karl Schwarzschild-Strasse 2, 85748 Garching bei Muenchen, Germany            
         \and
              Istituto Nazionale di Astrofisica, Osservatorio Astronomico di Roma, via Frascati 33, I-00040 Monte Porzio Catone, Italy
         \and
              South African Astronomical Observatory, PO Box 9, 7935 Observatory, South Africa
             }

   \date{Received May 14, 2008; accepted August 05, 2008}

% \abstract{}{}{}{}{} 
% 5 {} token are mandatory
 
  \abstract
  % context heading (optional)
  % {} leave it empty if necessary  
   {Classical Cepheids are excellent tracers of intermediate-mass stars, since their 
distances can be estimated with very high accuracy. In particular, they can be 
adopted to trace the chemical evolution of the Galactic disk.}
  % aims heading (mandatory)
   {Homogeneous iron abundance measurements for 33 Galactic Cepheids located in 
the outer disk together with accurate distance determinations based on near-infrared 
photometry are adopted to constrain the Galactic iron gradient beyond 10 kpc.}
  % methods heading (mandatory)
   {Iron abundances were determined using high resolution Cepheid spectra collected 
with three different observational instruments: ESPaDOnS@CFHT, Narval@TBL and 
FEROS@2.2m ESO/MPG telescope. Cepheid distances were estimated using near-infrared 
($J,H,K-$band) period-luminosity relations and data from SAAO and the 2MASS catalog.}
  % results heading (mandatory)
   {The least squares solution over the entire data set indicates that the iron gradient 
in the Galactic disk presents a slope of --0.052$\pm$0.003 \dexkpc{} in the 5-17 kpc range. 
However, the change of the iron abundance across the disk seems to be better 
described by a linear regime inside the solar circle and a flattening of 
the gradient toward the outer disk (beyond 10 kpc). In the latter region the 
iron gradient presents a shallower slope, i.e. --0.012$\pm$0.014 \dexkpc{}. 
In the outer disk (10-12 kpc) we also found that Cepheids present an 
increase in the spread in iron abundance. Current evidence indicates 
that the spread in metallicity depends on the Galactocentric longitude.  
Finally, current data do not support the hypothesis of a discontinuity 
in the iron gradient at Galactocentric distances of 10-12 kpc.}
  % conclusions heading (optional), leave it empty if necessary 
   {The occurrence of a spread in iron abundance as a function of the 
Galactocentric longitude indicates that linear radial gradients should 
be cautiously treated to constrain the chemical evolution across the disk.}


   \maketitle
%
%________________________________________________________________

\section{Introduction}

Chemodynamical evolutionary models provide interesting predictions 
concerning the formation and evolution of the Milky Way. However,
the plausibility and accuracy of these predictions need to be 
validated with observational data. The most commonly adopted 
observables are the star formation rate and the abundance 
gradients across the Galactic disk.

Different tracers have been used to determine the Galactic 
gradients : HII regions, Open Clusters, $O/B-$type stars and Cepheids.
Compared to other tracers, Cepheids present several advantages : 
{\it i)} they are well-known primary distance indicators; 
{\it ii)} they are bright enough to allow the study of the gradient 
over a large range of Galactocentric distances;
{\it iii)} they present a large set of well defined absorption lines, 
therefore, accurate abundances of many elements can be provided. 

Abundance gradients were discovered first by \citet{Sea71} in six 
external galaxies using HII regions.
However, their occurrence in our Galaxy remained controversial as 
they are not easy to observe due to the fact we are embedded
in the Galactic disk. Indeed, some studies were in favor of Galactic 
gradients, like \citet{D'odo76} or \citet{Jan79} whereas others 
found no correlation between distance and metallicity \citep{Cle73, Jen75}.
Today, the existence of Galactic abundance gradients seems widely 
accepted, even if some recent studies still report no evidence of 
such gradients, but the empirical determinations of their shapes 
and slopes are strongly debated.
 
Considering a simple linear gradient the slopes for different elements 
can be explored with several stellar tracers. The reader interested 
in exhaustive reviews concerning Galactic gradients is referred 
to \citet{And02a} and \citet{Chia01}. By using HII regions, 
\citet{Vil96} found a slope of --$0.02$ \dexkpc, but other authors 
using the same tracers suggested slopes ranging from --$0.039$ 
\dexkpc{} \citep{Deh00}  to --$0.065$ \dexkpc{} \citep{Aff97}. 
On the other hand, the use of B-type stars gives slopes ranging 
from --$0.042$ \dexkpc{} \citep{Daf04} to --$0.07$~\dexkpc{} \citep{Gum98}. 
The slopes based on planetary nebulae range from --$0.05$ \dexkpc{} 
\citep{Cos04} to --$0.06$ \dexkpc{} \citep{Mac99} to the lack of a 
Galactic metallicity gradient \citep{Sta06}. 
The slopes based on old Open Clusters still show a large spread. 
By adopting a sample of 40 clusters distributed 
between the solar circle and $R_G\simeq14$ kpc, \citet{Frie02} 
found a slope of --$0.06$ \dexkpc. 
More recently, \citet{Car07} using new accurate metal abundances 
for five old open clusters located in the outer disk together 
with the sample adopted by \citet{Frie02} found a much shallower 
global iron gradient, namely --$0.018$ \dexkpc.  
The global iron slopes based on Cepheids are very homogeneous, 
dating back to the first estimates by \citet{Har81,Har84}, 
who found a slope of --$0.07$ \dexkpc; the more recent estimates 
provide slopes ranging from --$0.06$ \dexkpc{} 
\citep{And02c,Luck03,And04,Luck06} to --$0.07$ \dexkpc{} \citep{Lem07}, 
hereafter \citetalias{Lem07}.

As far as the shape of the gradient is concerned, current findings are  
even more controversial. In particular, the hypothesis of a linear 
gradient is still widely debated. Several investigations based on different 
stellar tracers --- HII regions, \citet{Vil96}; open clusters, 
\citet{Twa97},\citet{Car07}; Cepheids, \citet{And04}; planetary
nebulae, \citet{Cos04}--- indicate a flattening of the Galactic 
gradient beyond 10-12 kpc.  
This flattening is well reproduced by chemodynamical evolutionary 
models \citep{Ces07}. More recently, \citet{Yong06} brought forward 
a new feature: the flattening may occur with two different basement 
values, the first one at --0.5 dex, a value more metal-poor than 
previous studies, and the second one at --0.8 dex. According to the 
quoted authors, the latter one might imply the possibility of a 
merger event. On the other hand, independent studies do not show 
evidence of a flattening toward the outer disk, like \citet{Roll00} 
based on O$/$B stars or \citet{Deh00} based on HII regions.
Moreover, a discontinuity in the metallicity gradient at 
$R_G\sim10-12$ kpc has also been suggested by \citet{Twa97} 
according to open cluster photometry. Instead of a regular 
decrease from the inner to the outer disk, they indicate 
a 2 zone distribution with an abrupt discontinuity  of $\sim$ --0.2 dex 
at about 10-12 kpc and a shallower gradient inside each zone. 
This hypothesis was supported by \citet{And02c}, \citet{Luck03} 
and by \citet{And04}.
Such variety may be due to the fact that the different tracers 
adopted to estimate the gradient present only a handful
of objects toward the outer disk, i.e. at $R_G\ge12$ kpc.

In this paper, we investigate the shape of the gradient in the 
outer disk and focus our attention on the transition zone around 
10-12 kpc. The paper is organized as follows: in \S 2 we describe 
the data and the reduction strategy adopted to estimate the iron 
abundances and the distances. In this section we also discuss the 
uncertainties affecting both abundances and distances. In \S 3 we
discuss the new iron abundances and compare them with previous 
measurements. In \S 4, we present the new Galactic iron gradients 
and compare them with previous estimates based on different stellar 
tracers. \S 5 summarizes current findings.     

%________________________________________________________________

\section{Observations and data reduction.}

\subsection{Observations.}
High resolution spectroscopic observations were performed in 2006 with ESPaDOnS at the CFHT (27 stars) 
and in 2007 with FEROS \citep{Kau99} at the 2.2m ESO/MPG telescope in La Silla 
(6~stars).  Another target (BK Aur) already included in the ESPaDOnS sample was also 
observed in 2007 with Narval at the 2m Telescope Bernard Lyot (TBL) in 
Pic du Midi in Southwest France. 
ESPaDOnS has a resolving power of 81,000 in the spectroscopic ``star-only'' mode 
(only the light from the star goes through the instrument) 
and an has accessible wavelength range of 370-1050 nm. 
FEROS has a resolving power of 48,000 and an accessible wavelength range of 
370-920 nm, while Narval is a copy of ESPaDOnS adapted to the specifics 
of the TBL, whose resolving power is 75,000 in the spectroscopic ``star-only'' 
mode and an accessible wavelength range of 370-1050 nm.
The spectra were reduced either using the FEROS package within MIDAS or 
using the Libre-ESpRIT software \citep{Don97,Don06}.
Relevant information concerning the observations and the signal-to-noise 
(S/N) ratio at 600 nm are listed in Table \ref{obslog}.
Phases were calculated with periods and epochs from the GCVS
(General Catalogue of Variable Stars) described in \citet{Sam04}, except for 8 stars to take 
into account changes in the period. In the case of AO Aur, RZ Gem, SV Mon, UY Mon, RS Ori, 
GQ Ori, they are from \citet{BerIgn00} while in the case of AD Gem and CV Mon they are from \citet{Sza91}.

\begin {table*}[!ht]
\begin{center}
\caption {Observations: date, phase, spectrograph, exposure time and S/N ratio at 600 nm.} 
\label {obslog}					       
\begin {tabular}{lcccccr} 
\hline
\hline
Object   &   date   & Julian date & phase & spectrograph & exp. time &  S/N \\
         &          &      d      &       &              &      s    & (600nm)\\ 
\hline 	                  	      
AO Aur   & 14/02/06 & 2453781.910 & 0,632 &  ESPaDOnS    &      1500 &  89 \\
AX Aur   & 16/02/06 & 2453783.775 & 0.619 &  ESPaDOnS    &    2*2700 & 113 \\
BK Aur   & 14/02/06 & 2453781.974 & 0.150 &  ESPaDOnS    &       700 & 101 \\
         & 31/10/07 & 2454405.736 & 0.096 &   NARVAL     &      1800 & 101 \\
SY Aur   & 15/02/06 & 2453782.928 & 0.809 &  ESPaDOnS    &       500 & 125 \\
 Y Aur   & 16/02/06 & 2453783.911 & 0.983 &  ESPaDOnS    &       800 & 142 \\
YZ Aur   & 15/02/06 & 2453782.712 & 0.789 &  ESPaDOnS    &      1600 & 145 \\
AO CMa   & 31/03/07 & 2454191.099 & 0.241 &   FEROS      &      2700 & 107 \\
AA Gem   & 15/02/06 & 2453782.818 & 0.749 &  ESPaDOnS    &       900 & 128 \\
AD Gem   & 15/02/06 & 2453782.971 & 0.131 &  ESPaDOnS    &       900 & 126 \\
RZ Gem   & 15/02/06 & 2453782.804 & 0.768 &  ESPaDOnS    &       800 &  91 \\
BE Mon   & 15/02/06 & 2453782.886 & 0.409 &  ESPaDOnS    &      1600 & 120 \\
BV Mon   & 16/02/06 & 2453783.950 & 0.851 &  ESPaDOnS    &      1800 &  79 \\
CV Mon   & 16/02/06 & 2453783.928 & 0.084 &  ESPaDOnS    &      1500 & 152 \\
EK Mon   & 15/02/06 & 2453782.838 & 0.809 &  ESPaDOnS    &      1800 & 103 \\
SV Mon   & 15/02/06 & 2453782.955 & 0.663 &  ESPaDOnS    &       300 & 119 \\
TW Mon   & 31/03/07 & 2454191.074 & 0.664 &   FEROS      &    2*1800 &  80 \\
TX Mon   & 14/02/06 & 2453781.932 & 0.998 &  ESPaDOnS    &      1700 & 111 \\
TY Mon   & 15/02/06 & 2453782.737 & 0.537 &  ESPaDOnS    &      1800 &  72 \\
TZ Mon   & 14/02/06 & 2453781.889 & 0.896 &  ESPaDOnS    &      1400 & 106 \\
UY Mon   & 16/02/06 & 2453783.967 & 0.065 &  ESPaDOnS    &       600 & 107 \\
V495 Mon & 15/02/06 & 2453782.911 & 0.074 &  ESPaDOnS    &      1800 &  46 \\
V508 Mon & 14/02/06 & 2453781.954 & 0.101 &  ESPaDOnS    &      1700 & 130 \\
V510 Mon & 31/03/07 & 2454191.029 & 0.975 &   FEROS      &    2*1800 &  98 \\
WW Mon   & 16/02/06 & 2453783.844 & 0.884 &  ESPaDOnS    &    2*2700 & 104 \\
XX Mon   & 15/02/06 & 2453782.863 & 0.584 &  ESPaDOnS    &      1800 &  58 \\
CS Ori   & 15/02/06 & 2453782.761 & 0.637 &  ESPaDOnS    &      1800 &  78 \\
GQ Ori   & 15/02/06 & 2453782.961 & 0.836 &  ESPaDOnS    &       300 & 125 \\
RS Ori   & 15/02/06 & 2453782.981 & 0.709 &  ESPaDOnS    &       350 & 118 \\
HW Pup   & 29/03/07 & 2454189.081 & 0.906 &   FEROS      &    4*1800 &  93 \\
AV Tau   & 16/02/06 & 2453783.722 & 0.701 &  ESPaDOnS    &    2*1800 &  73 \\
ST Tau   & 15/02/06 & 2453782.949 & 0.696 &  ESPaDOnS    &       300 & 120 \\
DD Vel   & 26/03/07 & 2454186.590 & 0.329 &   FEROS      &    4*1800 & 127 \\
EZ Vel   & 26/03/07 & 2454186.495 & 0.106 &   FEROS      &    4*1800 & 155 \\
         & 30/03/07 & 2454190.204 & 0.214 &   FEROS      &    2*1800 & 129 \\
\hline
\end {tabular}
\end {center}					       
\end {table*}			

\subsection{Line list}

We started from the iron line list provided by \citet{Rom08} and used 
in \citetalias{Lem07}.  This list was extended by following the method 
devised by the quoted authors to match the wider spectral range of 
ESPaDOnS and Narval. The atomic properties (oscillator strength, 
excitation potential) listed in VALD (Vienna Atomic Lines Database)
were adopted for all the selected lines. 

\subsection{Equivalent widths}

Measurements of equivalent widths (EW) are made with a semi-interactive code 
based on genetic algorithms from \citet{Char95}
(see \citetalias{Lem07} for details).
The analysis takes into account only weak, non blended lines so that all of 
them can be fitted by a Gaussian. While strong lines are systematically discarded, weak and 
asymmetric lines are removed after visual inspection. No peculiar treatment was applied 
to weakly asymmetric lines, like the one proposed by \citet{And05}.
To ensure the reliability of our EWs, we compared them to EWs obtained by direct integration of the lines in the case of RZ Gem.
As no systematics were found between direct integration and Gaussian fit, we assumed the asymmetry was weak enough 
so that the lines could still be fitted by Gaussians.
The number of lines finally used is on average 80-130 lines 
for $Fe I$ and 15-25 lines for $Fe II$.

\subsection{Effective temperature estimates}

The determination of an accurate temperature is a key point in the abundance 
determination. Following the same approach adopted in \citetalias{Lem07}, 
the stellar effective temperatures, T$_{eff}$, were estimated using the 
method of line depth ratios described in \citet{KovGor00}. This approach 
has the additional advantage of being independent of 
interstellar reddening. \citet{KovGor00} proposed 32 analytical relations 
to derive T$_{eff}$ from line depth ratios. After measuring the line 
depths and calculating associated ratios, we obtained 32 temperatures
whose mean value gave us the stellar temperatures listed 
in Table \ref{atmparam}.
Recently \citet{Kov07} increased to 131 the set of relations for the 
determination of supergiant temperature. We double-checked our temperatures 
using these new relations (Kovtyukh, private communication)
and we found a very good agreement between the two independent estimates.

\subsection{Atmospheric parameters}

The surface gravity, $log g$, and the micro-turbulent velocity, v$_{t}$, 
were determined by imposing an ionization balance between $Fe I$ and 
$Fe II$ with the help of the curve of growth. Lines of the same element 
in different ionization states should give the same abundance value.
Iterations on $log g$ and v$_{t}$ are, therefore, repeated until 
$Fe I$ and $Fe II$ adjust to the same curve of growth. This tool 
also provides an independent check for the T$_{eff}$ value, 
since both high and low $\chi_{ex}$ values properly fit the curve 
of growth. Once abundances were measured, atmospheric parameters
are validated by checking that the $Fe I$ abundances depend neither 
on line strength nor on the excitation potential. The atmospheric 
parameters for the Cepheids in our sample are listed in Table \ref{atmparam}.

\begin {table}[!ht]
\begin{center}
\caption {Atmospheric parameters adopted to compute iron abundances.} 
\scriptsize 
\label {atmparam}					       
\begin {tabular}{lccccc} 
\hline
\hline
 Object   & phase &  T$_{eff}$   & $log g$ &   V$_{t}$   & [Fe/H] \\
          &       &      K       &  dex    & km s$^{-1}$ &  dex   \\
 \hline       	        					  
 AO Aur   & 0,632 &     5450     &   0.8   &    3.0      & -0.41  \\
 AX Aur   & 0.619 &     5635     &   0.8   &    2.7      & -0.22  \\
 BK Aur   & 0.150 &     5970     &   1.3   &    3.0      & -0.10  \\
          & 0.096 &     6160     &   1.4   &    3.4      & -0.04  \\
 SY Aur   & 0.809 &     6220     &   1.1   &    3.0      & -0.07  \\
  Y Aur   & 0.983 &     6300     &   1.5   &    3.0      & -0.26  \\
 YZ Aur   & 0.789 &     5715     &   1.0   &    3.5      & -0.33  \\
 AO CMa   & 0.241 &     6065     &   1.7   &    3.7      & -0.04  \\
 AA Gem   & 0.749 &     5500     &   1.0   &    4.7      & -0.35  \\
 AD Gem   & 0.131 &     6130     &   1.6   &    3.0      & -0.19  \\
 RZ Gem   & 0.768 &     5500     &   0.8   &    3.3      & -0.44  \\
 BE Mon   & 0.409 &     5865     &   1.4   &    3.0      & -0.07  \\
 BV Mon   & 0.851 &     6080     &   1.9   &    3.2      & -0.10  \\
 CV Mon   & 0.084 &     6350     &   2.0   &    3.1      & -0.10  \\
 EK Mon   & 0.809 &     5700     &   1.3   &    3.4      & -0.05  \\
 SV Mon   & 0.663 &     4900     &   0.5   &    3.4      & -0.10  \\
 TW Mon   & 0.664 &     5640     &   1.3   &    3.4      & -0.15  \\
 TX Mon   & 0.998 &     5945     &   1.0   &    3.0      & -0.12  \\
 TY Mon   & 0.537 &     5555     &   1.2   &    3.0      & -0.15  \\
 TZ Mon   & 0.896 &     6020     &   1.3   &    3.0      & -0.04  \\
 UY Mon   & 0.065 &     6100     &   1.5   &    3.0      & -0.33  \\
 V495 Mon & 0.074 &     6090     &   1.6   &    3.2      & -0.17  \\
 V508 Mon & 0.101 &     6100     &   1.5   &    3.2      & -0.25  \\
 V510 Mon & 0.975 &     5390     &   0.8   &    3.3      & -0.12  \\
 WW Mon   & 0.884 &     6650     &   1.4   &    3.0      & -0.32  \\
 XX Mon   & 0.584 &     5520     &   0.9   &    3.0      & -0.18  \\
 CS Ori   & 0.637 &     6350     &   2.0   &    3.1      & -0.19  \\
 GQ Ori   & 0.836 &     4960     &   2.6   &    0.8      &  0.11  \\
 RS Ori   & 0.709 &     5475     &   1.3   &    3.0      & -0.14  \\
 HW Pup   & 0.906 &     5300     &   0.6   &    3.1      & -0.28  \\
 AV Tau   & 0.701 &     5750     &   1.3   &    3.0      & -0.17  \\
 ST Tau   & 0.696 &     5700     &   1.4   &    3.2      & -0.14  \\
 DD Vel   & 0.329 &     5600     &   1.1   &    3.2      & -0.35  \\
 EZ Vel   & 0.106 &     6380     &   1.5   &    3.6      & -0.01  \\
          & 0.214 &     5455     &   0.6   &    3.6      & -0.01  \\
\hline
\end {tabular}
\end {center}					       
\end {table}	

\subsection{Abundance determinations}

Atmospheric parameters are then used as input for stellar atmosphere models.
We used the MARCS models of \citet{Edv93} which are based on the following assumptions:
plane-parallel stratification, hydrostatic equilibrium and LTE.
The abundance determination codes adjust abundances until a good match 
between predicted and observed equivalent widths is obtained.
The final abundance of a star is estimated as the mean value of 
the abundances determined for each line.
The high S/N ratio of the spectra and the use of a substantial number 
of clean lines provide the opportunity to reduce the intrinsic errors 
on iron abundance to $\sim 0.12$ dex. Note that in this investigation, 
we adopted the solar chemical abundances provided by \citet{Gre96}.

\subsection{Uncertainties in abundance determination}

The very first source of errors in abundance determination comes from 
the extraction of data from the spectra.
The equivalent width (EW) and the continuum placement must be accurately 
measured. Thus, we limited our analysis to symmetric and unblended lines, 
avoiding the most crowded parts of the spectra. We retained only relatively 
weak lines (EW $\leq$ 120 m\AA) for the abundance calculation and checked 
that they could be fitted by a Gaussian.

The crucial point in the abundance determination is to obtain an accurate temperature as it strongly affects the
line strength. As mentioned above, we used the method of line depth ratios from \citet{KovGor00}, according to which
the uncertainties on absolute temperatures are at most $\approx$~80 K. Moreover, the curves of growth provided an 
independent check of the temperature determination as lines with high and low $\chi_{ex}$ values must both fit well
the curve of growth. This ensures us that our intrinsic dispersion in temperature is not more than 100 K.

In order to test how the uncertainties affect the final abundance result, we computed abundances by adopting
over or under-estimated values of the atmospheric parameters.
As expected, only the errors on temperature have a noticeable effect: over or under-estimating the 
effective temperature of 100~K implies a difference of about $\pm$ 0.1 dex in the abundance determination.
Errors on the surface gravity of $\Delta log g= \pm0.3$ dex and on the micro-turbulent velocity of 
$\Delta v_{t} = \pm 0.5$ km/sec have modest effects as the differences are only $\pm$ 0.03 and $\pm$ 0.05~dex respectively.
The sum in quadrature of these uncertainties on the atmospheric parameters gives uncertainties on the abundances of $\sim$ 0.12 dex. 

Finally, the LTE assumption might not be appropriate for supergiant, variable stars. The adopted stellar
atmosphere models could then lead to systematic errors. However, the use of relatively weak lines should 
minimize this effect. Moreover, \citet{FryCar97} have shown that a canonical spectroscopic analysis, using LTE
atmosphere models, gives reliable abundances for Cepheids : dwarfs and Cepheids located in the same open clusters
are found to have the same metallicity, within the errors. \citet{Yong06} also tested the influence of NLTE
effects and showed that the surface gravities derived from a classical approach are robust.

\subsection{Distances}

Individual Cepheid distances were estimated using near-infrared ($J,H,K-$band) 
photometry either collected from SAAO or from the 2MASS catalog. 
The mean magnitude of the Cepheids based on 2MASS photometry was estimated 
using the template light curves provided by \citet{Sos05} together with the 
$V-$band amplitude and the epoch of maximum available in the literature. 
The data from \citet{Bar97} were also used for ST~Tau and AD~Gem while for 
HW~Pup, the data from \citet{Sche92} were used together with 2MASS data. 
In the case of UY~Mon and AO~CMa that are first overtone pulsators, it is 
not possible to use the empirical template, since it is  only available 
for fundamental mode pulsators. The mean near-infrared magnitude
of these objects is only based on the 2MASS measurement. Their period was
fundamentalized as explained below.
Note that the SAAO magnitudes were transformed into the 2MASS photometric 
system using the transformations provided by \citet{Koen07}.

We adopted the individual Cepheid reddenings provided either by \citet{Lan07} 
or listed in the Fernie database \citep{Fer95}. These two sets of reddenings 
are not homogeneous, as recent results provided by \citet{Lan07} are corrected 
for metallicity effects whereas reddenings from \citet{Fer95} do not take into account 
metallicity effects. The relative absorption in the $V-$band 
was estimated using A$_{V}$ = 3.1 E(B-V) with A$_{J}$ = 0.28 A$_{V}$, A$_{H}$ = 0.19 A$_{V}$, 
and A$_{K}$ = 0.11 A$_{V}$ \citep{Car89}. 

The Cepheid distances were estimated using the near-infrared Period-Luminosity 
(PL) relations provided by \citet{Per04} and by assuming an LMC distance 
modulus of 18.50, a conservative value adopted by the HST Key Project \citep{Free01}.
For a discussion on the LMC distance modulus, please see \citet{Rom08} 
and references therein. Note that the quoted PL relations were also transformed 
into the 2MASS photometric system using the transformations provided by 
\citet{Car01}. In order to estimate the distance of first overtone Cepheids 
their periods were fundamentalized, i.e. we added 0.127 to their logarithmic period.

Using these PL relations from \citet{Per04} for Galactic Cepheids is justified 
although they are based on a large and homogeneous sample of LMC Cepheids.
Current empirical and theoretical findings indicate that the slope and 
the zero-point of NIR PL relations are not strongly dependent on metallicity 
\citep{Bono99,Per04,Fou07,Rom08}. Moreover, the near-infrared PL relations 
present an intrinsic dispersion that is significantly smaller than the 
dispersion of the optical PL relations. The difference is due to the fact 
that the former ones are, at fixed period, marginally affected by the 
intrinsic width in temperature of the instability strip. Finally, Cepheid 
distances based on near-infrared PL relations are more accurate than optical 
ones because they are much less affected by uncertainties in reddening 
estimates.   

The heliocentric distance of the stars adopted in the following is the mean 
value between the three distances in $J$, $H$, and $K$ bands to limit random 
errors. The Galactocentric distance is calculated with the classical 
formula assuming a Galactocentric distance of 8.5 kpc for the sun \citep{Fea97},
in order to enable comparisons with \citetalias{Lem07}.
This determination (whose error bars are $\pm$0.5 kpc) was based on the analysis 
of the Hipparcos proper motion of 220 Cepheids. Recent studies give Galactocentric 
distances for the sun between 7.62$\pm$0.32 kpc \citep{Eis05} and 8.8$\pm$0.3 kpc
\citep{Coll06}. For a recent review of the distance to the Galactic Center, see \citet{Groe08}.

The distances based on near-infrared photometry are given in the second column 
of Table \ref{distances}. Galactocentric distances based on $V-$band photometry
for which we adopted the heliocentric distances given in the Fernie database 
are given {\it only as a guideline} in the first column (see \citet{Fer95} for details). 
Indeed these distances were estimated using a $V-$band period-luminosity relation but this relation 
is not corrected for metallicity effects and was derived using individual 
reddenings determined from a period-colour relation which doesn't account for the metallicity,
whereas these period-colour relations are likely to be metallicity-dependent \citep{LS94,Bono99}.\\ 
As expected, the two different sets of distance determinations agree quite well 
near the solar circle, where the metallicities are in general near to the solar value.
However, for lower metallicity objects located in the outer disk, 
the discrepancy increases and reaches respectively 2.8\%, 3.5\%, 3.4\% and 4.8\% 
in the case of V510 Mon, TW Mon, HW Pup and YZ Aur, 4 stars among the 5 more distant 
Cepheids in our sample.

\begin {table}[!ht]
\begin {center}
\caption[]{Cepheid distances: \\\hspace{\linewidth} 1st column: Galactocentric distance from V band photometry. 
\\\hspace{\linewidth} 2nd column: Galactocentric distance from IR photometry.
\\\hspace{\linewidth} 3rd column: Heliocentric distance from IR photometry.}
\label {distances}					       
\begin {tabular}{lccc} 
\hline
\hline
  Object & Galactocentric & Galactocentric & Heliocentric \\
         &  distance (V)  &  distance (IR) & distance (IR)\\
         &      kpc       &      kpc       &     kpc      \\        
 \hline                	           
 AO Aur  &     12.60      &     12.43      &     3.940    \\
 AX Aur  &     12.79      &     12.75      &     4.256    \\
 BK Aur  &     10.72      &     10.86      &     2.500    \\
 SY Aur  &     10.71      &     10.92      &     2.486    \\
  Y Aur  &     10.27      &     10.42      &     1.969    \\
 YZ Aur  &     13.30      &     12.71      &     4.278    \\
 AO CMa  &     11.32      &     11.10      &     4.068    \\
 AA Gem  &     12.39      &     12.31      &     3.820    \\
 AD Gem  &     11.20      &     11.37      &     2.939    \\
 RZ Gem  &     10.57      &     10.66      &     2.158    \\
 BE Mon  &     10.02      &     10.11      &     1.742    \\
 BV Mon  &     10.89      &     11.19      &     2.986    \\
 CV Mon  &     10.22      &     10.09      &     1.774    \\
 EK Mon  &     10.84      &     10.71      &     2.575    \\
 SV Mon  &     10.94      &     10.82      &     2.514    \\
 TW Mon  &     14,47      &     14.00      &     6.079    \\
 TX Mon  &     12.56      &     12.34      &     4.344    \\
 TY Mon  &     11.90      &     11.94      &     3.890    \\
 TZ Mon  &     12.22      &     12.01      &     3.982    \\
 UY Mon  &     10.74      &     10.37      &     2.038    \\
 V495 Mon&     12.79      &     12.65      &     4.679    \\
 V508 Mon&     11.38      &     11.43      &     3.222    \\
 V510 Mon&     13.73      &     13.38      &     5.337    \\
 WW Mon  &     13.68      &     13.52      &     5.274    \\
 XX Mon  &     12.68      &     12.55      &     4.617    \\
 CS Ori  &     12.42      &     12.56      &     4.214    \\
 GQ Ori  &     10.91      &     10.83      &     2.469    \\
 RS Ori  &     10.00      &     10.02      &     1.575    \\
 HW Pup  &     13.28      &     13.77      &     7.795    \\
 AV Tau  &     11.51      &     11.54      &     3.043    \\
 ST Tau  &      9.50      &      9.52      &     1.044    \\
 DD Vel  &       -        &     10.69      &     6.710    \\
 EZ Vel  &       -        &     12.20      &     9.522    \\
\hline
\end {tabular} 
\end {center}
\end {table}		

%________________________________________________________________

\section{Results : Iron abundances}

A large fraction of Cepheids in our sample present slightly sub-solar iron abundances, i. e. the metal content expected for Cepheids
located in the outer disk. Current results are in very good agreement with previous studies, mainly the series of papers from 
Andrievsky and collaborators \citep{And02a,And02b,And02c,Luck03,And04,And05,Kov05a,Kov05b,Luck06}. 
Indeed 17 stars present small ($<$ 0.1 dex) differences in iron abundances. On the other hand, for six stars the differences 
are $\sim$ 0.1 dex and only for four stars (UY Mon, BK Aur, RZ Gem and AO Aur) is the difference higher and approaches 0.3 dex.

The differences could arise from the methods employed to determine abundances, as Andrievsky and collaborators used
a modified version of the standard spectroscopic analysis in order to avoid a possible dependence of $Fe I$ lines on NLTE effects, 
as explained in \citet{KovAnd99}. 
In their method, surface gravities and microturbulent velocities are derived from $Fe II$ lines instead of $Fe I$ lines. However, the differences are not 
correlated with specific phases, atmospheric parameters or metallicities. Another difference between the studies is the value of atomic parameters, as \citet{KovAnd99}
derived their $log g$ through an inverted analysis of the solar spectrum. However, $log g$ for $Fe I$ are very similar for lines in common. Our list of $Fe II$ lines 
is about twice as large due to the wider spectral range, and there are only mild differencies for lines in common, except for 2 lines where they reach $\approx$ 0.2 dex. 
Thus differences in atomic parameters cannot be the origin of the discrepancies in iron abundances. 

\begin{figure*}[!ht]
	\sidecaption
	\includegraphics[angle=-90, width = 12 cm]{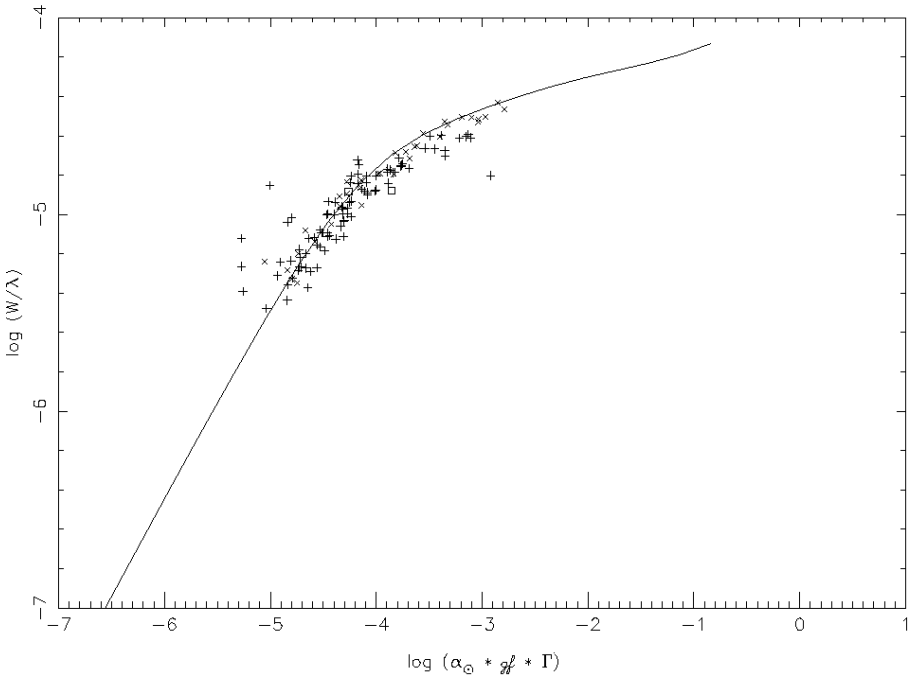}
	\caption{ Observed curve of growth for $Fe I$ in RZ Gem. The full line is the theoretical curve of growth for a typical line ($\lambda$=5000\AA, $\chi_{ex}$=3).
The atmospheric parameters adopted for this star are T$_{eff}$=5500 K, log g=0.8, v$_{t}$=3.3 km/s, and [Fe/H]=-0.44 dex. 
Squares represent lines with $\chi_{ex}$ $<$ 1.5 crosses lines with 1.5 $<$  $\chi_{ex}$ $<$ 3.0 and pluses lines with $\chi_{ex}$ $>$ 3.0.}
        \label{FeI_RZGem}
\end{figure*}

\begin{figure*}[!ht]
	\sidecaption
	\includegraphics[angle=-90, width = 12 cm]{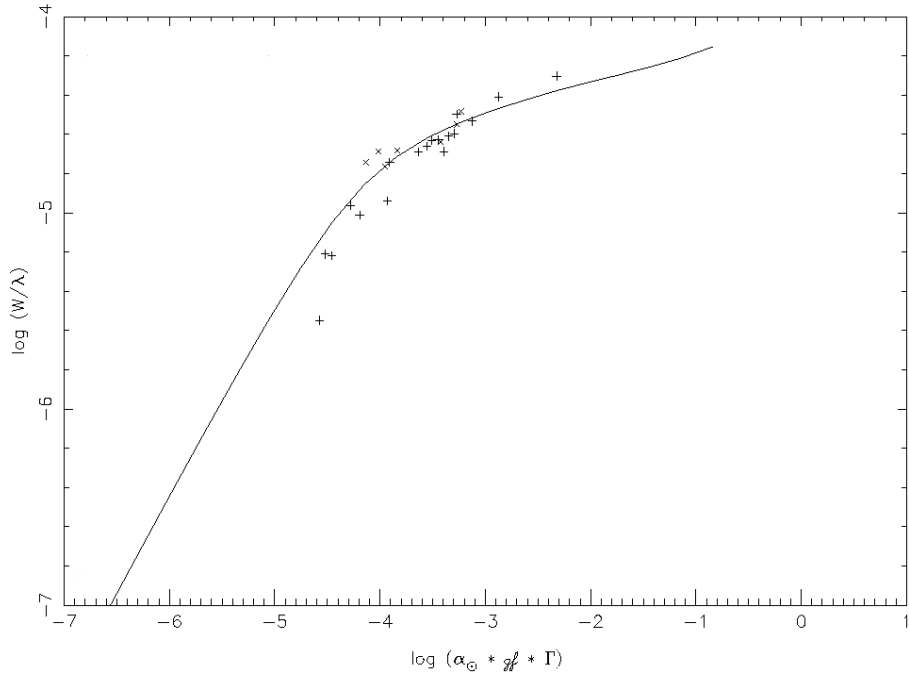}
	\caption{ Same as Fig. \ref{FeI_RZGem} but for $Fe II$.}
        \label{FeII_RZGem}
\end{figure*}

The curves of growth for $Fe I$ (Fig. \ref{FeI_RZGem}) and for $Fe II$ (Fig. \ref{FeII_RZGem}) are presented below in the case of RZ Gem.
The good agreement between predicted curves of growth and measurements for both $Fe I$ and $Fe II$ as well as the test performed on EWs measurements 
(see \S 2.3) indicate that our abundances measurements are robust. 
The same outcome applies for the other discrepant Cepheids. However, these stars deserve further investigation as they are all located in the outer disk,
which is less well sampled than the solar neighborhood.

%%% Curve of growth RZ Gem : FeI

%%% Curve of growth RZ Gem : FeII

A few stars have been investigated in many studies and the results are quite often homogeneous (see, e. g., SV Mon). 
Only three stars of our sample (YZ Aur, WW Mon and HW Pup) are in common with both Andrievsky's series and \citet{Yong06}. 
Although our classical approach is very similar to the analysis performed by \citet{Yong06}, current iron abundances agree quite well with those published in
\citet{Luck06}. We found metal abundances that are higher than the values provided by \citet{Yong06}. 

This difference cannot be explained by the different approaches adopted in the two investigations to estimate the effective temperature. 
Indeed, independent support of current T$_{eff}$ values is provided by the evidence that both high and low $\chi_{ex}$ lines properly fit the curve of growth.  
Moreover, we found that the two investigations adopted different $gf$-values for the $Fe II$ lines, but the difference is on average smaller than 0.1 dex
and the ensuing difference in the estimated surface gravities is negligible. Therefore, this cannot account for the differences in iron
abundances. Although our spectra have higher spectral resolution (48000-81000 versus 28000) and higher S/N ratio
(145, 104, 93 versus 49, 60, 49) than the spectra from\citet{Yong06}, the discrepancies correlate neither with the  S/N ratio nor with
the atmospheric parameters. We cannot reach any firm conclusion regarding this discrepancy, since there are only three objects in common in the two Cepheid samples.

\begin {table}[!ht]
\begin {center}
\caption[]{Cepheid iron abundance and comparisons with previous results: 
\\\hspace{\linewidth} (1) \citet{And02a}, (2) \citet{And02c}, 
\\\hspace{\linewidth} (3) \citet{Luck03}, (4) \citet{And04},
\\\hspace{\linewidth} (5) \citet{Luck06}, (6) \citet{Yong06},
\\\hspace{\linewidth} (7) \citet{Kov05a}, (8) \citet{Kov05b},
\\\hspace{\linewidth} (9) \citet{Pont97}.
}
\label {iron}					       
\begin {tabular}{lcc} 
\hline
\hline
 Object   &   [Fe/H]  & [Fe/H] : Previous results \\
          &     dex   &          dex              \\
 \hline 
 AO Aur   &  -0.41    & -0.14$^{4}$               \\
 AX Aur   &  -0.22    &                           \\
 BK Aur   &  -0.10    &  0.17$^{5}$               \\
          &  -0.04    &                           \\
 SY Aur   &  -0.07    & -0.02$^{8}$               \\
  Y Aur   &  -0.26    & -0.23$^{5}$               \\
 YZ Aur   &  -0.33    & -0.07$^{1}$, -0.37$^{5}$, -0.64$^{6}$, -0.36$^{8}$ \\
 AO CMa   &  -0.04    & -0.02$^{9}$                     \\
 AA Gem   &  -0.35    & -0.24$^{4}$               \\
 AD Gem   &  -0.19    & -0.19$^{4}$               \\
 RZ Gem   &  -0.44    & -0.12$^{4}$               \\
 BE Mon   &  -0.07    &                           \\
 BV Mon   &  -0.10    &                           \\
 CV Mon   &  -0.10    & -0.03$^{1}$               \\
 EK Mon   &  -0.05    & -0.1$^{2}$                \\
 SV Mon   &  -0.10    & -0.03$^{1}$, 0.00$^{2}$, -0.04$^{8}$, -0.02$^{7}$ \\
 TW Mon   &  -0.15    & -0.24$^{3}$, -0.20$^{9}$  \\
 TX Mon   &  -0.12    & -0.14$^{2}$               \\
 TY Mon   &  -0.15    &                           \\
 TZ Mon   &  -0.04    & -0.03$^{2}$, -0.12$^{3}$, -0.24$^{9}$ \\
 UY Mon   &  -0.33    & -0.08$^{4}$               \\
 V495 Mon &  -0.17    & -0.26$^{2}$, -0.16$^{9}$  \\
 V508 Mon &  -0.25    & -0.25$^{2}$               \\
 V510 Mon &  -0.12    & -0.19$^{3}$, -0.20$^{6}$  \\
 WW Mon   &  -0.32    & -0.29$^{3}$, -0.25$^{9}$, -0.59$^{6}$ \\
 XX Mon   &  -0.18    & -0.18$^{2}$, -0.08$^{3}$  \\
 CS Ori   &  -0.19    & -0.26$^{2}$               \\
 GQ Ori   &   0.11    & -0.03$^{1}$, 0.06$^{5}$   \\
 RS Ori   &  -0.14    & -0.1$^{1}$,-0.02$^{3}$    \\
 HW Pup   &  -0.28    & -0.29$^{2}$, -0.2$^{3}$, -0.40$^{6}$ \\
 AV Tau   &  -0.17    &                           \\
 ST Tau   &  -0.14    & -0.05$^{1}$               \\
 DD Vel   &  -0.35    &                           \\
 EZ Vel   &  -0.01    &                           \\
\hline		       
\end {tabular}					       
\end {center}
\end {table}

%________________________________________________________________ 

\section{Results : Galactic abundance gradients}

\subsection{The iron gradient along the whole Galactic disk}

\begin{figure*}[!ht]
	\sidecaption
	\includegraphics[angle=90,width=12cm]{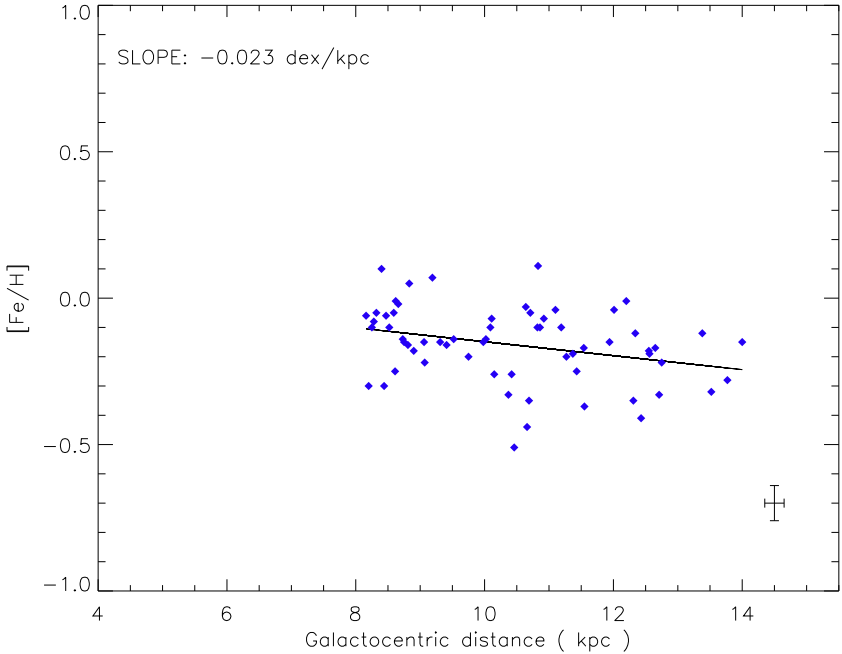}
\caption{Galactic iron abundance gradient. The solid line shows the linear regression over the entire sample. Characteristic error bars
are plotted in the lower right corner.}
\label{fig:04-15_our_data}
\end{figure*}

The current sample includes 33 Cepheids and we added to this new data the 30 Cepheids of our previous investigation \citep{Lem07}.
The 63 stars of this sample are located between 8 and 15 kpc from the Galactic center (see Fig. \ref{fig:04-15_our_data}).
We investigated at first the working hypothesis of a linear gradient and calculated the iron Galactic gradient using the entire Cepheid sample.
We found a slope of --0.023$\pm$0.007 \dexkpc{} that is shallower than previous studies. However, this low value can be easily explained
by the lack of targets inside the Solar circle : in this region, the metallicities are higher and indeed, inner Cepheids approach abundances
of the order of +0.3 dex \citep{And02b}. The inclusion of these Cepheids will then dramatically change the slope of the gradient.
Moreover and even more importantly, data plotted in Fig. \ref{fig:04-15_our_data} do not show evidence of a gap over the Galactocentric distances 
covered by the current Cepheid sample.

%%% gradient : 04--15_our_data

\begin{figure*}[!h]
	\sidecaption
	\includegraphics[width=12cm]{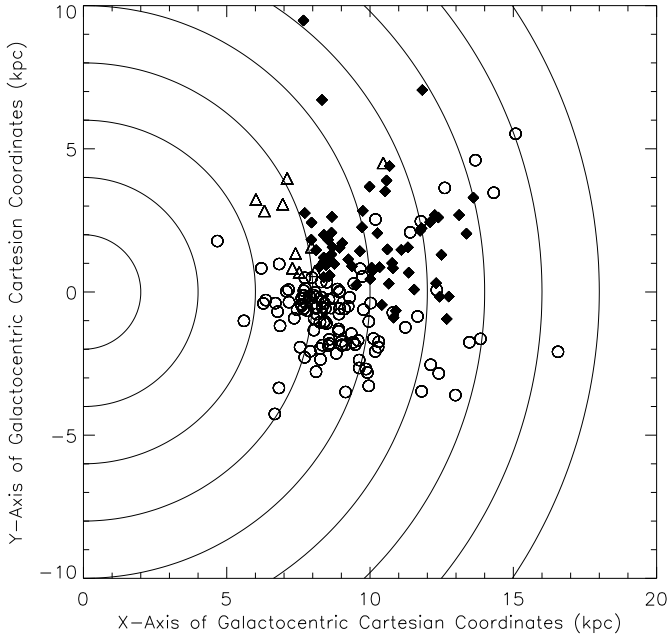}
\caption{Spatial distribution of the entire Cepheid sample. Filled diamonds mark our data, while open circles show data from Andrievsky et al. 
and open triangles are data from Mottini et al. The x-axis contains both the center of the Galaxy at (0,0) and the Sun at (8.5,0). 
Galactocentric circles with radii of 2 to 18 kpc are overplotted on the map.}
\label{fig:polrep}
\end{figure*}

In order to improve the sampling across the Galactic disk, we added to these 63 Cepheids 115 stars from Andrievsky's sample 
and 10 stars from \citet{Rom08}. For the entire Cepheid sample, we have both metallicities and infrared photometry. 
Among them, 38 Cepheids have Galactocentric distances between 10 and 12 kpc and 27 are even more distant.
Individual distances and iron abundances are listed in Table \ref{addstars}. A map (as seen from above the Galaxy) 
showing the spatial distribution of the Cepheids is displayed in Fig. \ref{fig:polrep}. The x-axis contains both
the center of the Galaxy at (0,0) and the Sun at (8.5,0). Galactocentric circles with radii of 2 to 18 kpc are overplotted on the map.

%%% Polar repartition 

\begin {table*}[!t]
\begin {center}
\caption[] {Additional stars from Andrievsky's series and from \citet{Rom08}. The latter are marked with an (*).
\\\hspace{\linewidth} Distances are based on infrared photometry and [Fe/H] found in the literature.
} 
\scriptsize 
\begin {tabular}{lrrlrrlrrlrrlrr} 
\hline
\hline
 Object   & d$_{IR}$ & [Fe/H] &  Object   & d$_{IR}$ & [Fe/H] &  Object   & d$_{IR}$ & [Fe/H] &  Object    & d$_{IR}$ & [Fe/H] &  Object   & d$_{IR}$ & [Fe/H] \\
          &    kpc   &   dex  &           &    kpc   &   dex  &           &    kpc   &   dex  &            &    kpc   &   dex  &           &    kpc   &   dex  \\
 \hline 	
   TT Aql &   7.70   &   0.11 &    BD Cas &   9.85   &  -0.07 &    MW Cyg &   8.15   &   0.09 &     S Nor &   7.74   &   0.05  &   RY Sco   &  7.16 &  0.09 \\       	       
   FM Aql &   7.86   &   0.08 &    CF Cas &  10.30   &  -0.01 &  V386 Cyg &   8.48   &   0.11 &  V340 Nor &   6.90   &   0.00  &   KQ Sco   &  6.26 &  0.16 \\              
   FN Aql &   7.41   &  -0.02 &    CH Cas &   9.98   &   0.17 &  V402 Cyg &   8.18   &   0.02 &     Y Oph &   7.94   &   0.05  & V500 Sco   &  7.08 & -0.02 \\                
 V496 Aql &   7.63   &   0.05 &    CY Cas &  10.48   &   0.06 &  V532 Cyg &   8.55   &  -0.04 &    BF Oph &   7.69   &   0.00  &   SS Sct   &  7.61 &  0.06 \\                
V1162 Aql &   7.48   &   0.01 &    DD Cas &  10.22   &   0.10 &  V924 Cyg &   7.92   &  -0.09 &    SV Per &  10.82   &   0.01  &   UZ Sct   &  5.69 &  0.33 \\      
  Eta Aql &   8.30   &   0.05 &    DF Cas &  10.45   &   0.13 & V1334 Cyg &   8.46   &  -0.04 &    UX Per &  12.30   &  -0.21  &   EW Sct   &  8.14 &  0.04 \\      
 V340 Ara &   5.00   &   0.31 &    DL Cas &   9.46   &  -0.01 & V1726 Cyg &   8.73   &  -0.02 &    VX Per &  10.46   &  -0.05  & V367 Sct   &  6.81 & -0.01 \\      
   RT Aur &   8.93   &   0.06 &    FM Cas &   9.58   &  -0.09 &  Beta Dor &   8.50   &  -0.01 &    AS Per &   9.79   &   0.10  &   BQ Ser   &  7.70 & -0.04 \\      
   RX Aur &  10.03   &  -0.07 &  V636 Cas &   8.96   &   0.06 &     W Gem &   9.44   &  -0.04 &    AW Per &   9.29   &   0.01  &   SZ Tau   &  8.94 &  0.08 \\      
   AN Aur &  11.69   &  -0.16 &     V Cen &   7.99   &   0.04 &  Zeta Gem &   8.86   &   0.04 &    BM Per &  11.30   &   0.10  &   AE Tau   & 12.33 & -0.19 \\      
   CY Aur &  13.58   &  -0.40 &    CP Cep &   9.79   &  -0.01 &     V Lac &   9.17   &   0.00 &    HQ Per &  13.95   &  -0.31  &    S Vul   &  7.60 & -0.02 \\      
   ER Aur &  16.69   &  -0.34 &    CR Cep &   9.00   &  -0.06 &     X Lac &   9.01   &  -0.02 &  V440 Per &   9.13   &  -0.05  &    T Vul   &  8.35 &  0.01 \\      
   RW Cam &  10.01   &   0.04 &    IR Cep &   8.65   &   0.11 &     Y Lac &   9.08   &  -0.09 &     X Pup &  10.50   &  -0.03  &    U Vul   &  8.16 &  0.05 \\      
   RX Cam &   9.25   &   0.03 &   Del Cep &   8.57   &   0.06 &     Z Lac &   9.20   &   0.01 &     S Sge &   8.13   &   0.10  &    X Vul   &  8.10 &  0.08 \\      
   TV Cam &  12.38   &  -0.08 &    BG Cru &   8.30   &  -0.02 &    RR Lac &   9.21   &   0.13 &     U Sgr &   7.89   &   0.04  &   SV Vul   &  7.80 &  0.03 \\      
   AB Cam &  12.72   &  -0.09 &     X Cyg &   8.32   &   0.12 &    BG Lac &   8.77   &  -0.01 &     W Sgr &   8.08   &  -0.01  &    U Car   &  8.11 &  0.12$^{*}$ \\
   AD Cam &  13.47   &  -0.22 &    SU Cyg &   8.17   &   0.00 &  V473 Lyr &   8.33   &  -0.06 &     Y Sgr &   8.02   &   0.06  &   WZ Car   &  8.14 &  0.18$^{*}$ \\
   RW Cas &  10.39   &   0.22 &    SZ Cyg &   8.60   &   0.09 &     T Mon &   9.75   &   0.22 &    VY Sgr &   6.29   &   0.26  &   VW Cen   &  6.83 &  0.05$^{*}$ \\
   RY Cas &   9.91   &   0.26 &    TX Cyg &   8.46   &   0.20 &    AA Mon &  12.03   &  -0.21 &    WZ Sgr &   6.71   &   0.17  &   XX Cen   &  7.52 &  0.04$^{*}$ \\
   SU Cas &   8.71   &  -0.01 &    VX Cyg &   8.58   &   0.09 &    CU Mon &  14.73   &  -0.26 &    XX Sgr &   7.18   &   0.10  &   KN Cen   &  6.92 &  0.07$^{*}$ \\
   SW Cas &   9.34   &   0.13 &    VY Cyg &   8.49   &   0.00 &    EE Mon &  16.06   &  -0.51 &    AP Sgr &   7.68   &   0.10  &   GH Lup   &  7.56 &  0.03$^{*}$ \\
   SY Cas &   9.66   &   0.04 &    VZ Cyg &   8.75   &   0.05 &    FG Mon &  14.43   &  -0.20 &    AV Sgr &   6.37   &   0.34  &    S Mus   &  8.12 &  0.18$^{*}$ \\
   SZ Cas &  10.25   &   0.04 &    BZ Cyg &   8.54   &   0.19 &    FI Mon &  13.12   &  -0.18 &    BB Sgr &   7.71   &   0.08  &   UU Mus   &  7.60 &  0.05$^{*}$ \\
   TU Cas &   8.95   &   0.03 &    CD Cyg &   8.05   &   0.07 &  V504 Mon &  11.59   &  -0.31 &  V350 Sgr &   7.62   &   0.18  &    U Nor   &  7.34 &  0.10$^{*}$ \\
   XY Cas &   9.73   &   0.03 &    DT Cyg &   8.42   &   0.11 &  V526 Mon &   9.69   &  -0.13 &    RV Sco &   7.74   &   0.10  &   LS Pup   & 11.39 & -0.11$^{*}$ \\
\hline  
\end {tabular}					       
\end {center}
\label {addstars}					       
\end {table*}

\begin{figure*}[!h]
	\sidecaption
	\includegraphics[angle=90, width=12cm]{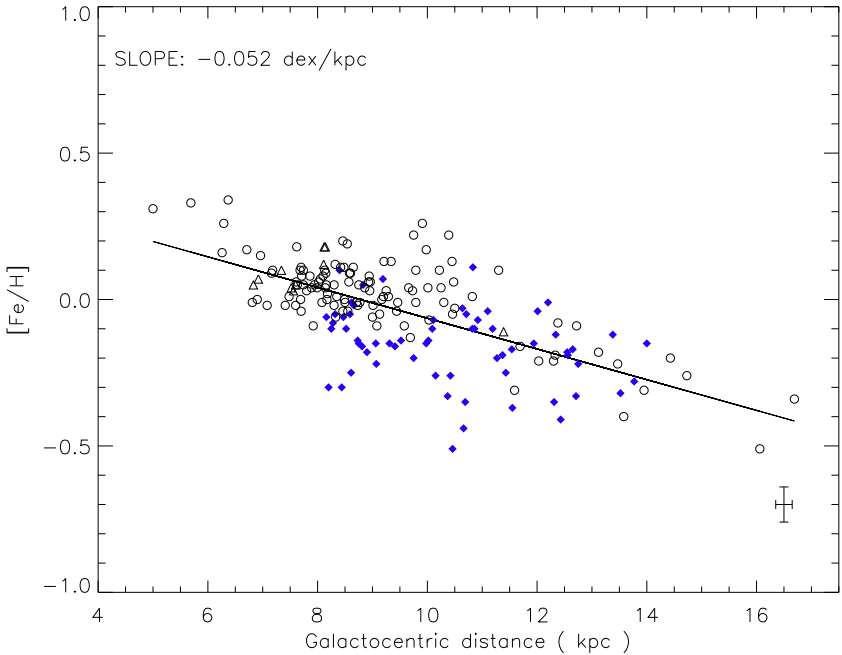}
\caption{Galactic iron abundance gradient. Filled diamonds mark our data, open circles show data from Andrievsky et al. 
and open triangles data from Mottini et al. The solid line shows the linear regression over the entire sample. 
Characteristic error bars are plotted in the lower right corner.}
\label{fig:04-17_full_data}
\end{figure*}

By adopting the entire sample we calculated once again the iron Galactic gradient. Note that the new sample includes 60 Cepheids 
located inside the Solar circle together with two Cepheids located beyond 16 kpc.
We found that the slope of the linear iron gradient over the whole sample is --0.052$\pm$0.003 \dexkpc{} (Fig. \ref{fig:04-17_full_data}).
This slope is slightly shallower than in \citetalias{Lem07} where we found --0.061$\pm$0.019 \dexkpc, as well as in \citet{Frie02} 
who found --0.06$\pm$0.01 \dexkpc{} from open clusters and in \citet{Luck06} who found --0.068$\pm$0.003 \dexkpc{} from classical Cepheids.
Data plotted in this figure indicate that:
\begin{itemize}
	\item The spread in metallicity is larger in the outer disk, in particular around 10 kpc.
	\item There is no evidence of a gap in the Galactocentric distances covered by the current sample. We will discuss
this point in the next section.
\end{itemize}

%%% gradient : 04--17_full_data 

\subsection{The iron gradient in the outer disk}

\begin{figure*}[!h]
	\sidecaption
	\includegraphics[angle=90,width=12cm]{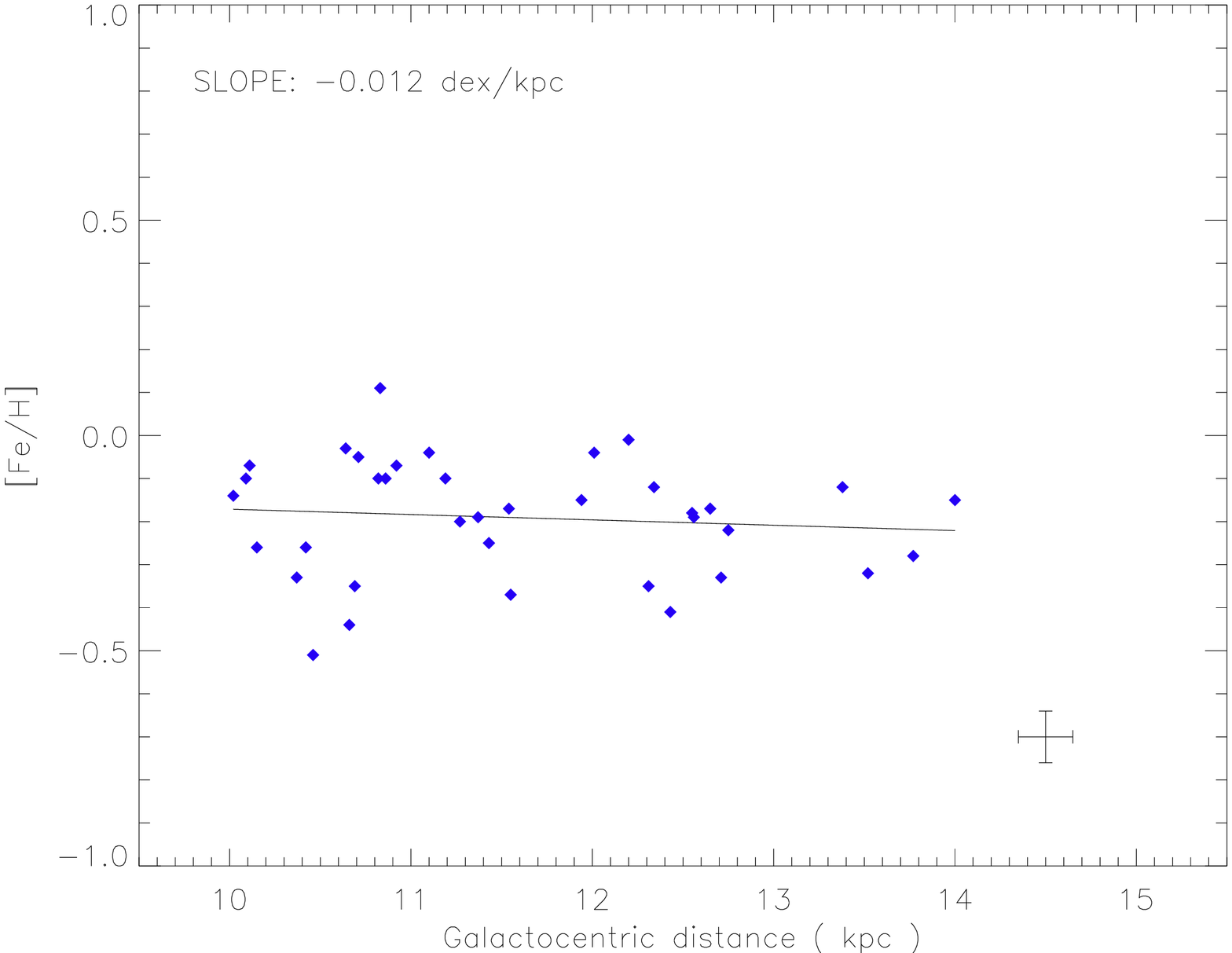}
\caption{Same as Fig. \ref{fig:04-15_our_data}, but for Cepheids with Galactocentric distances ranging from 10 to 15 kpc.}
\label{fig:10-15_our_data}
\end{figure*}

To properly investigate the iron gradient in the outer disk, we selected (in our sample of 63 stars) Cepheids located between 10 and 15 kpc
(see Fig. \ref{fig:10-15_our_data}). We estimated the Galactic gradient and found a slope of --0.012$\pm$0.014 \dexkpc{}. 
The shallow slope clearly indicates a flattening of the iron gradient in the outer disk. This also suggests that the Galactic iron gradient
might be more accurately described by a bimodal distribution. This evidence agrees quite well with the slope (--0.018$\pm$0.02 \dexkpc) recently 
found by \citet{Car07} using old Open Clusters ranging from 12 to 21 kpc and a Galactocentric distance of the  Sun of 8.5 kpc. 
It is also similar to the flat distribution (+0.004$\pm$0.011) found by \citet{And04} in the outer zone (zone III) of their multimodal gradient.  
Moreover and even more importantly, the two quoted independent investigations suggest, within the errors, a similar result concerning the value
of the metallicity basement. The value of the metallicity basement we found from our data is --0.19$\pm$0.14 dex, whereas \citet{Car07} found that open clusters
located in the outer disk have an iron abundance of [Fe/H] = --0.34$\pm$0.15 dex, a value close to the $\sim$ --0.3 dex found by \citet{Twa97} from the same kind
of objects. Our basement value also agrees very well with the one of \citet{And04} as the mean iron abundance obtained in their outer disk zone is --0.209$\pm$0.058 dex.  

%%% gradient : 10--15_our_data 

Current findings are marginally consistent with the results concerning the metallicity basement by \citet{Yong06}. They suggested 
a two-level basement in the outer disk: one level located at $\sim$ --0.5 dex, is notably lower than the other values, 
while the second one is still lower at $\sim$ --0.8 dex. According to these authors, the more metal--poor values might be explained by Cepheids acquired 
by the Galaxy through merging events. This working hypothesis was also supported by the significantly different [$\alpha$/$Fe$] ratios shown by these objects.
In this context, \citet{Yong05} and \citet{Car05} using different stellar tracers, namely open clusters and field red giants, found
a metallicity basement of $\approx$ --0.5 dex. 

\begin{figure*}[!h]
	\sidecaption
	\includegraphics[angle=90, width=12cm]{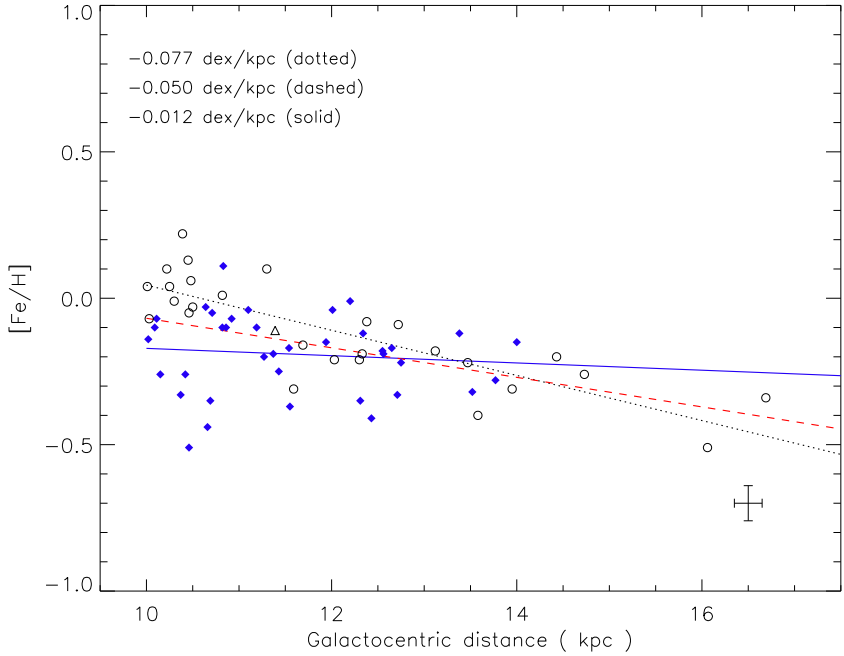}
\caption{Galactic iron abundance gradient. Filled diamonds mark our data, while open circles show data from Andrievsky et al. 
and open triangles display data from Mottini et al. The solid line shows the slope based on our sample, the dotted line the slope based
on the additional data from Andrievsky et al. and the dashed line the slope obtained using the two quoted samples. 
Characteristic error bars are plotted in the lower right corner.}
\label{fig:10-15_data}
\end{figure*}

If we now study the same region including all the Cepheids for which we have accurate estimates of both iron abundance and distance, 
the gradient is affected by selection effects.
The value of the slope depends on:
\begin{itemize}
	\item the range of Galactocentric distances covered by the Cepheid sample, since the targets are not uniformly distributed along the disk. 
	\item the spatial sampling across the disk. By using Cepheids in our sample with $R \geq$ 10 kpc we found a slope of- -0.012$\pm$0.014 \dexkpc{}. 
On the other hand, if we use Cepheids from the Andrievsky sample and our distances we found a slope of --0.077$\pm$0.010 \dexkpc{}, while
with two samples together we obtain a slope of --0.050$\pm$0.008 \dexkpc{} (See Fig. \ref{fig:10-15_data}). 
\end{itemize}

%%% gradient : 10--15_data 

The quoted results are strongly affected by the poor spatial sampling. The number of known Cepheids located in the outer disk is 
quite limited and quite often they are also relatively faint. This means that long exposure times with 4-8m class telescopes are 
necessary to obtain the S/N required for accurate spectroscopy.

The other interesting findings brought forward by the data plotted in Fig. \ref{fig:10-15_data} is the large spread in metallicity 
between 10 and 12 kpc. This feature was partially expected and is caused by the decrease in gas density and in star formation rate when moving from
the innermost to the outermost regions of the Galaxy.
\begin{itemize}
	\item A group of Cepheids has rather high metallicities ($\sim$ +0.0 to $\sim$ +0.2 dex). Some of them were already 
noted by \citet{Luck06} to be located approximately at the same Galactic longitude $\sim$ 120 \deg.
	\item On the contrary, a group of stars has rather low metallicities ($\sim$ --0.2 to $\sim$ --0.4 dex), most of them
lying in the 180\deg $<$ $l$ $<$ 200\deg range of Galactic longitudes.
\end{itemize}

\begin{figure*}[!h]
	\sidecaption
	\includegraphics[width=12cm]{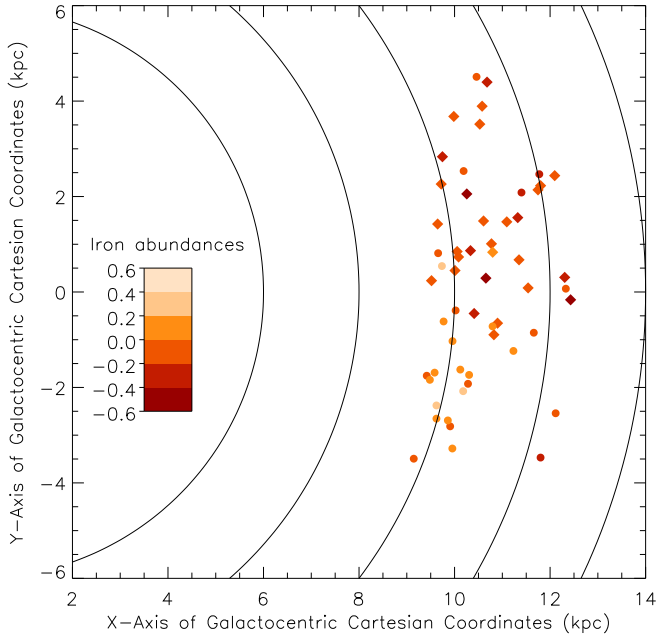}
\caption{Metallicity of Cepheids located between 10 and 12 kpc. Filled diamonds mark our data while filled circles show data from Andrievsky et al.
The metallicity strongly depends on the Galactocentric longitude (see text). The center of the Galaxy is at (0,0) and the Sun is at (8.5,0).
Galactocentric circles with radii of respectively at 6, 8, 10, 12 and 14 kpc are over-plotted on the map.}
\label{fig:met_map_10}
\end{figure*}

According to this circumstantial empirical evidence, if we plot the Cepheids located between 10 and 12 kpc on a Galactocentric polar map 
(see Fig. \ref{fig:met_map_10}), we can see that the metallicity is dependent on the Galactocentric longitude, with a spread in 
metallicity reaching 0.8 dex for stars located at roughly the same Galactocentric distance. Here we define the Galactocentric longitude as the angle of a polar 
coordinate system centered on the Galactic center. Galactocentric longitudes are measured from --180\deg to $+$180\deg, 
counter clockwise from the origin, arbitrarily defined as the Galactic radius containing the Galactic center at (0,0) and the Sun at (8.5,0).
Distances projected on Galactocentric Cartesian coordinates are also labelled on the x and y-axis.

%%% gradient : polar map of the metallicity around 10 kpc 

The occurrence of a large spread in metal abundance between stars located at the same radius of the Galactic disk allows us to explain why we do not 
observe a gap in the Galactic gradient as proposed in some previous studies. The gap might be caused either by selection effects in the sampling 
of the disk, or by local inhomogeneities in the disk, and therefore, restricted to a narrow range in longitude. The Cepheid sample collected by
\citet{And02c,Luck03,And04} are indeed located in a limited range of Galactic longitude (190\deg $<$ $l$ $<$ 250\deg). 
The same outcome applies to the sample of 76 Open Clusters collected by \citet{Twa97} (130\deg $<$ $l$ $<$ 260\deg).

Current findings suggest the need to be very cautious in using the slopes of Galactic gradients, in particular in the outer disk due to poor spatial sampling.
A significant improvement in the sample of outer disk Cepheids requires a systematic photometric survey to detect new Cepheids. This goal can be achieved 
with the new generation of automatic telescopes such as HAT (Hungarian-made Automated Telescope; http://cfa-www.harvard.edu/$\sim$gbakos/HAT/index.html) or 
ASAS (All Sky Automated Survey; http://archive.princeton.edu/$\sim$asas/) and with spectroscopic follow up of the new targets. 

%________________________________________________________________

\section{Conclusions}
High resolution spectra and accurate distance determinations based on infrared photometry allowed us to extend our study of the Galactic
iron abundance gradient toward the outer disk. New data led us to a slope of --0.052$\pm$0.003 \dexkpc{} in the 5-17 kpc range, in very 
good agreement with previous studies if we consider a linear gradient. However, the Galactic iron abundance gradient is better described 
by a bimodal distribution with a steeper slope toward the bulge and a flattening of the gradient toward the outer disk. 
In particular, we found a shallow slope of --0.012$\pm$0.014 \dexkpc{} in this region (10-15 kpc). 
Current data show no evidence of a sharp discontinuity in metallicity for Galactocentric radii of $\sim$~10-12 kpc. 
Poor spatial sampling affecting the adopted stellar tracers do not allow us to constrain on a quantitative basis whether the occurrence of this gap is real or caused by local
inhomogeneities in the metallicity distribution of the disk. We also found a large spread in metal abundance ($\approx$ 0.8 dex) for Cepheids located approximately
at the same distance (10 $\leq R \leq$ 12 kpc), but covering a broad range of Galactocentric longitude.
A larger Cepheid sample, uniformly distributed over the four Galactic quadrants, is mandatory to constrain this effect.

%________________________________________________________________

\begin{acknowledgements}
This research has made use of the SIMBAD and VIZIER database, operated at CDS, Strasbourg, France.
We thank the staff of the CFHT and of the TBL and SciOps at La Silla Observatory for their support during observations.
We deeply thank V. V. Kovtyukh for checking our temperatures with his new relations for line depth ratios and S. M. Andrievsky, our referee,
for his very useful comments and suggestions.
\end{acknowledgements}

\end{document}